\documentclass[12pt,a4paper]{article}
\makeatletter
\def\eqnarray{%
\stepcounter{equation}%
\let\@currentlabel=\theequation
\global\@eqnswtrue
\global\@eqcnt\z@
\tabskip\@centering
\let\\=\@eqncr
$$\halign to \displaywidth\bgroup\@eqnsel\hskip\@centering
$\displaystyle\tabskip\z@{##}$&\global\@eqcnt\@ne
\hfil$\displaystyle{{}##{}}$\hfil
&\global\@eqcnt\tw@$\displaystyle\tabskip\z@{##}$\hfil
\tabskip\@centering&\llap{##}\tabskip\z@\cr}
\makeatother
\newcommand{\ket}[1]{{\vert{#1}\rangle}}
\newcommand{\bra}[1]{{\langle{#1}\vert}}
\newcommand{\kett}[1]{{\vert{#1}\rangle\rangle}}
\newcommand{\braa}[1]{{\langle\langle{#1}\vert}}
\newcommand{\calh}{{\cal H}}

\newcommand{\cala}{{\cal A}}

\newcommand{\fukuso}{{\mathbf C}}
\newcommand{\futon}{{\bf N}}

\newcommand{\zetta}{{\vert z\vert}}

\newcommand{\kappazetta}{{\vert\kappa\vert}}

\begin{document}

\title{\sl Mathematical Structure of Rabi Oscillations \\  
           in the Strong Coupling Regime}
\author{
  Kazuyuki FUJII
  \thanks{E-mail address : fujii@yokohama-cu.ac.jp }\\
  Department of Mathematical Sciences\\
  Yokohama City University\\
  Yokohama, 236-0027\\
  Japan
  }
\date{}
\maketitle\thispagestyle{empty}
%
%
%
%
\begin{abstract}
  In this paper we generalize the Jaynes--Cummings Hamiltonian 
  by making use of some operators based on Lie algebras su(1,1) and su(2), 
  and study a mathematical structure of Rabi floppings of these models 
  in the strong coupling regime. We show that Rabi frequencies are given 
  by matrix elements of generalized coherent operators (quant--ph/0202081) 
  under the rotating--wave approximation. 
  
  In the first half we make a general review of coherent operators and 
  generalized coherent ones based on Lie algebras su(1,1) and su(2). 
  In the latter half we carry out a detailed examination of 
  Frasca (quant--ph/0111134) and generalize his method, 
  and moreover present some related problems.  
  
  We also apply our results to the construction of controlled unitary gates 
  in Quantum Computation. 
  Lastly we make a brief comment on application to Holonomic Quantum 
  Computation. 
\end{abstract}

\newpage

%
%
%
%

\section{Introduction}

Coherent states or generalized coherent states play an important role in 
quantum physics, in particular, quantum optics, see \cite{KS} and \cite{MW}. 
They also play an important one in mathematical physics. See 
the textbook \cite{AP}. For example, they are very useful in performing 
stationary phase approximations to path integral, \cite{FKSF1}, 
\cite{FKSF2}, \cite{FKS}. 

Coherent operators which produce coherent states are very useful because 
they are unitary and easy to handle. 
The basic reason is probably that they are subject to the elementary 
Baker-Campbell-Hausdorff (BCH) formula. Many basic properties of them 
are well--known, see \cite{AP} or \cite{KF6}. 

Generalized coherent operators which produce generalized coherent states 
are also useful. But they are not so easy to handle in spite of having  
the disentangling one corresponding to the elementary BCH formula. 
In \cite{KF5} and \cite{KF14} the author determined all matrix elements 
of generalized coherent operators based on Lie algebras su(1,1) and su(2). 
They are interesting by themselves, but moreover have a very interesting 
application. 

In \cite{MFr} Frasca dealt with the Jaynes--Cummings model which 
describes a two--level atom interacting with a single radiation mode 
(see \cite{MSIII} for a general review) in the strong coupling regime 
(not weak coupling one !) and showed that Rabi frequencies are obtained by 
matrix elements of coherent operator under the rotating--wave approximation. 
His aim was to explain the recent experimental finding on Josephson 
junctions \cite{NPT}. 

This is an interesting result and moreover his method can be widely 
generalized. See also \cite{MFr2} for an another example dealt with in the 
strong coupling regime. 

In this paper we generalize the Jaynes--Cummings Hamiltonian 
by making use of some operators based on Lie algebras su(1,1) and su(2), 
and study a mathematical structure of Rabi floppings of these extended 
models in the strong coupling regime. 

We show that (generalized) Rabi frequencies are also given by 
matrix elements of generalized coherent operators under the rotating--wave 
approximation. 
We believe that the results will give a new aspect to Quantum Optics or 
Mathematical Physics. 

We also apply our results to the construction of controlled unitary gates 
in Quantum Computation in the last section. 

Lastly we discuss an application to Holonomic Quantum Computation, but 
our discussion is not complete.

\section{Coherent and Generalized Coherent Operators}

\subsection{Coherent Operator}
Let $a(a^\dagger)$ be the annihilation (creation) operator of the harmonic 
oscillator.
If we set $N\equiv a^\dagger a$ (:\ number operator), then
\begin{equation}
  \label{eq:2-1-1}
  [N,a^\dagger]=a^\dagger\ ,\
  [N,a]=-a\ ,\
  [a^\dagger, a]=-\mathbf{1}\ .
\end{equation}
Let $\calh$ be a Fock space generated by $a$ and $a^\dagger$, and
$\{\ket{n}\vert\  n\in\futon\cup\{0\}\}$ be its basis.
The actions of $a$ and $a^\dagger$ on $\calh$ are given by
\begin{equation}
  \label{eq:2-1-2}
  a\ket{n} = \sqrt{n}\ket{n-1}\ ,\
  a^{\dagger}\ket{n} = \sqrt{n+1}\ket{n+1}\ ,
  N\ket{n} = n\ket{n}
\end{equation}
where $\ket{0}$ is a normalized vacuum ($a\ket{0}=0\  {\rm and}\  
\langle{0}\vert{0}\rangle = 1$). From (\ref{eq:2-1-2})
state $\ket{n}$ for $n \geq 1$ are given by
\begin{equation}
  \label{eq:2-1-3}
  \ket{n} = \frac{(a^{\dagger})^{n}}{\sqrt{n!}}\ket{0}\ .
\end{equation}
These states satisfy the orthogonality and completeness conditions
\begin{equation}
  \label{eq:2-1-4}
   \langle{m}\vert{n}\rangle = \delta_{mn}\ ,\quad \sum_{n=0}^{\infty}
   \ket{n}\bra{n} = \mathbf{1}\ . 
\end{equation}

\noindent{\bfseries Definition}\quad We call a state  
\begin{equation}
\label{eq:2-1-7}
\ket{z} =  \mbox{e}^{za^{\dagger}- \bar{z}a}\ket{0}\equiv U(z)\ket{0} 
\quad \mbox{for}\quad z\ \in\ \fukuso 
\end{equation}
the coherent state.

\subsection{Generalized Coherent Operator Based on $su(1,1)$}
Let us state generalized coherent operators and states based on $su(1,1)$.

We consider a spin $K\ (> 0)$ representation of $su(1,1) 
\subset sl(2,\fukuso)$ and set its generators 
$\{ K_{+}, K_{-}, K_{3} \}\ ((K_{+})^{\dagger} = K_{-})$, 
\begin{equation}
  \label{eq:2-2-3}
 [K_{3}, K_{+}]=K_{+}, \quad [K_{3}, K_{-}]=-K_{-}, 
 \quad [K_{+}, K_{-}]=-2K_{3}.
\end{equation}
We note that this (unitary) representation is necessarily infinite 
dimensional. 
The Fock space on which $\{ K_{+}, K_{-}, K_{3} \}$ act is 
$\calh_K \equiv \{\ket{K,n} \vert n\in\futon\cup\{0\} \}$ and 
whose actions are
\begin{eqnarray}
  \label{eq:2-2-4}
 K_{+} \ket{K,n} &=&  \sqrt{(n+1)(2K+n)}\ket{K,n+1},\quad 
 K_{-} \ket{K,n}  =  \sqrt{n(2K+n-1)}\ket{K,n-1},         \nonumber \\
 K_{3} \ket{K,n} &=& (K+n)\ket{K,n}, 
\end{eqnarray}
where $\ket{K,0}$ is a normalized vacuum ($K_{-}\ket{K,0}=0$ and 
$\langle K,0|K,0 \rangle =1$). We have written $\ket{K,0}$ instead 
of $\ket{0}$  to emphasize the spin $K$ representation, see \cite{FKSF1}. 
From (\ref{eq:2-2-4}), states $\ket{K,n}$ are given by 
\begin{equation}
  \label{eq:2-2-5}
 \ket{K,n} =\frac{(K_{+})^n}{\sqrt{n!(2K)_n}}\ket{K,0} ,
\end{equation}
where $(a)_n$ is the Pochammer's notation \ 
$
 (a)_n \equiv  a(a+1) \cdots (a+n-1).
$
These states satisfy the orthogonality and completeness conditions 
\begin{equation}
  \label{eq:2-2-7}
  \langle K,m \vert K,n \rangle =\delta_{mn}, 
 \quad \sum_{n=0}^{\infty}\ket{K,n}\bra{K,n}\ = \mathbf{1}_K.
\end{equation}
Now let us consider a generalized version of coherent states : 

\noindent{\bfseries Definition}\quad We call a state  
\begin{equation}
   \label{eq:2-2-8}
 \ket{z} = V(z)\ket{K,0} \equiv  
 \mbox{e}^{zK_{+} - \bar{z}K_{-}} \ket{K,0}  
  \quad \mbox{for} \quad z \in \fukuso.
\end{equation}
the generalized coherent state (or the coherent state of Perelomov's 
type based on $su(1,1)$ in our terminology).

Here let us construct an example of this representation. 
First we set
\begin{equation}
  \label{eq:2-2-12}
  K_{+}\equiv{1\over2}\left(a^{\dagger}\right)^2\ ,\
  K_{-}\equiv{1\over2}a^2\ ,\
  K_{3}\equiv{1\over2}\left(a^{\dagger}a+{1\over2}\right)\ ,
\end{equation}
then it is easy to check that these satisfy the commutation relations 
(\ref{eq:2-2-3}). 
That is, the set $\{K_{+},K_{-},K_{3}\}$ gives a unitary representation of 
$su(1,1)$with spin $K = 1/4\ \mbox{and}\ 3/4$, \cite{AP}. 
Now we also call an operator 
\begin{equation}
  \label{eq:2-2-14}
   S(z) = \mbox{e}^{\frac{1}{2}\{z(a^{\dagger})^2 - \bar{z}a^2\}}
   \quad \mbox{for} \quad z \in \fukuso 
\end{equation}
the squeezed operator, see the book \cite{AP}.

\subsection{Generalized Coherent Operator Based on $su(2)$}
Let us state generalized coherent operators and states based on $su(2)$.

We consider a spin $J\ (> 0)$ representation of $su(2) 
\subset sl(2,\fukuso)$ and set its generators 
$\{ J_{+}, J_{-}, J_{3} \}\ ((J_{+})^{\dagger} = J_{-})$, 
\begin{equation}
  \label{eq:2-3-3}
 [J_{3}, J_{+}]=J_{+}, \quad [J_{3}, J_{-}]=-J_{-}, 
 \quad [J_{+}, J_{-}]=2J_{3}.
\end{equation}
We note that this (unitary) representation is necessarily finite 
dimensional. 
The Fock space on which $\{ J_{+}, J_{-}, J_{3} \}$ act is 
$\calh_{J} \equiv \{\ket{J,n} \vert 0 \le n \le 2J \}$ and 
whose actions are
\begin{eqnarray}
  \label{eq:2-3-4}
 J_{+} \ket{J,n} &=& \sqrt{(n+1)(2J-n)}\ket{J,n+1}, \quad 
 J_{-} \ket{J,n}  =  \sqrt{n(2J-n+1)}\ket{J,n-1},        \nonumber \\
 J_{3} \ket{J,n} &=& (-J+n)\ket{J,n}, 
\end{eqnarray}
where $\ket{J,0}$ is a normalized vacuum ($J_{-}\ket{J,0}=0$ and 
$\langle J,0|J,0 \rangle =1$). We have written $\ket{J,0}$ instead 
of $\ket{0}$ to emphasize the spin $J$ representation, see \cite{FKSF1}. 
From (\ref{eq:2-3-4}), states $\ket{J,n}$ are given by 
\begin{equation}
  \label{eq:2-3-5}
 \ket{J,n} =\frac{(J_{+})^n}{\sqrt{n!{}_{2J}P_n}}\ket{J,0}.
\end{equation}
These states satisfy the orthogonality and completeness conditions 
\begin{equation}
  \label{eq:2-3-6}
  \langle J,m \vert J,n \rangle =\delta_{mn}, 
  \quad \sum_{n=0}^{2J}\ket{J,n}\bra{J,n}\ = \mathbf{1}_{J}.
\end{equation}
Now let us consider a generalized version of coherent states : 

\noindent{\bfseries Definition}\quad We call a state  
\begin{equation}
   \label{eq:2-3-7}
 \ket{z} = W(z)\ket{J,0} \equiv 
 \mbox{e}^{zJ_{+} - \bar{z}J_{-}} \ket{J,0}  
  \quad \mbox{for} \quad z \in \fukuso.
\end{equation}
the generalized coherent state (or the coherent state of Perelomov's 
type based on $su(2)$ in our terminology).

\par \vspace{3mm} \noindent
A comment is in order. 
We can construct the spin $K$ and $J$ representations by making 
use of Schwinger's boson method. But we don't repeat here, see for 
example \cite{KF5}.

\section{Matrix Elements of Coherent and Generalized Coherent Operators 
$\cdots$ \cite{KF14} }

\subsection{Matrix Elements of Coherent Operator}

We list matrix elements of coherent operators $U(z)$. 

\noindent{\bfseries The Matrix Elements}\quad The matrix elements of 
$U(z)$  are :
\begin{eqnarray}
   \label{eq:3-1-1-1}
 &&(\mbox{i})\quad n \le m \quad 
   \bra{n}U(z)\ket{m} = \mbox{e}^{-\frac{1}{2}\zetta^2}\sqrt{\frac{n!}{m!}}
                 (-\bar{z})^{m-n}{L_n}^{(m-n)}(\zetta^2), \\
   \label{eq:3-1-1-2}
 &&(\mbox{ii})\quad n \geq m \quad 
   \bra{n}U(z)\ket{m} = \mbox{e}^{-\frac{1}{2}\zetta^2}\sqrt{\frac{m!}{n!}}
                 z^{n-m}{L_m}^{(n-m)}(\zetta^2),
\end{eqnarray}
where ${L_n}^{(\alpha)}$ is the associated Laguerre's polynomial defined by 
\begin{equation}
   \label{eq:3-1-2}
 {L_k}^{(\alpha)}(x)=\sum_{j=0}^{k}(-1)^j {{k+\alpha}\choose{k-j}}
                  \frac{x^j}{j!}. 
\end{equation}
In particular $L_{k}\equiv {L_k}^{(0)}$ is the usual Laguerre's polynomial 
and these are related to diagonal elements of $U(z)$.

\subsection{Matrix Elements of Coherent Operator Based on $su(1,1)$}

We list matrix elements of $V(z)$  coherent operators based on $su(1,1)$. 
In this case it is always $2K > 1$ ($2K=1$ under some regularization). 

\noindent{\bfseries The Matrix Elements}\quad The matrix elements of 
$V(z)$ are :
\begin{eqnarray}
   \label{eq:3-2-1-1}
  (\mbox{i})\quad n \le m \quad 
   && \bra{K,n}V(z)\ket{K,m}= 
    \sqrt{\frac{n!m!}{(2K)_n(2K)_m}}
    {(-\bar{\kappa})^{m-n}}(1+\kappazetta^2)^{-K-\frac{n+m}{2}}
    \ \times   \nonumber \\
   &&\sum_{j=0}^{n}(-1)^{n-j}\frac{\Gamma(2K+m+n-j)}{\Gamma(2K)(m-j)!
     (n-j)!j!}(1+\kappazetta^2)^j(\kappazetta^2)^{n-j}, \\ 
   \label{eq:3-2-1-2}
  (\mbox{ii})\quad n \geq m \quad    
    && \bra{K,n}V(z)\ket{K,m}=
    \sqrt{\frac{n!m!}{(2K)_n(2K)_m}}
    {\kappa^{n-m}}(1+\kappazetta^2)^{-K-\frac{n+m}{2}}
    \ \times   \nonumber \\
   &&\sum_{j=0}^{m}(-1)^{m-j}\frac{\Gamma(2K+m+n-j)}{\Gamma(2K)(m-j)!
     (n-j)!j!}(1+\kappazetta^2)^j(\kappazetta^2)^{m-j}, 
\end{eqnarray}
where 
\begin{equation}
   \label{eq:3-2-2}
\kappa \equiv \frac{\mbox{sinh}(\zetta)}{\zetta}z 
       ={\mbox{cosh}(\zetta)}\zeta.
\end{equation}

The author doesn't know whether or not the right hand sides of 
(\ref{eq:3-2-1-1}) and (\ref{eq:3-2-1-2}) could be written by 
making use of some special functions such as generalized Laguerre's 
functions in (\ref{eq:3-1-2}). Therefore we set temporarily 
\begin{equation}
{F_{m}}^{(n-m)}(x:2K)=
     \sum_{j=0}^{m}(-1)^{m-j}\frac{\Gamma(2K+m+n-j)}{\Gamma(2K)(m-j)!
     (n-j)!j!}(1+x)^j{x}^{m-j} 
\end{equation}
and ${F_{m}}^{(0)}(x;2K)=F_{m}(x;2K)$.

\subsection{Matrix Elements of Coherent Operator Based on $su(2)$}

We list matrix elements of $W(z)$  coherent operators based on $su(2)$.
In this case it is always $2J \in \futon$.

\noindent{\bfseries  Matrix Elements}\quad The matrix elements of 
$W(z)$ are :
\begin{eqnarray}
   \label{eq:3-3-1-1}
  (\mbox{i})\quad n \le m \quad 
   && \bra{J,n}W(z)\ket{J,m}= 
    \sqrt{\frac{n!m!}{{}_{2J}P_n {}_{2J}P_m}}
    (-\bar{\kappa})^{m-n}(1-\kappazetta^2)^{J-\frac{n+m}{2}}
    \ \times   \nonumber \\
   &&\sum_{j=0}^{n}{}_{*}(-1)^{n-j}\frac{(2J)!}{(2J-m-n+j)!(m-j)!(n-j)!j!}
     (1-\kappazetta^2)^{j}(\kappazetta^2)^{n-j},\quad    \\ 
   \label{eq:3-3-1-2}
  (\mbox{ii})\quad n \geq m \quad    
    && \bra{J,n}W(z)\ket{J,m}=
    \sqrt{\frac{n!m!}{{}_{2J}P_n {}_{2J}P_m}}
    \kappa^{n-m}(1-\kappazetta^2)^{J-\frac{n+m}{2}}
    \ \times   \nonumber \\
   &&\sum_{j=0}^{m}{}_{*}(-1)^{m-j}\frac{(2J)!}{(2J-m-n+j)!(m-j)!(n-j)!j!}
     (1-\kappazetta^2)^{j}(\kappazetta^2)^{m-j},\quad  
\end{eqnarray}
where 
\begin{equation}
   \label{eq:3-3-2}
\kappa \equiv \frac{\mbox{sin}(\zetta)}{\zetta}z 
       ={\mbox{cos}(\zetta)}\eta.
\end{equation}
Here $\sum{}_{*}$ means a summation over $j$ satisfying $2J-m-n+j \geq 0$. 

The author doesn't know whether or not the right hand sides of 
(\ref{eq:3-3-1-1}) and (\ref{eq:3-3-1-2}) could be written 
by making use of some special functions. We set temporarily
\begin{equation}
{F_{m}}^{(n-m)}(x:2J)=
     \sum_{j=0}^{m}{}_{*}(-1)^{m-j}\frac{(2J)!}{(2J-m-n+j)!(m-j)!(n-j)!j!}
     (1-x)^{j}{x}^{m-j}
\end{equation}
and ${F_{m}}^{(0)}(x;2J)=F_{m}(x;2J)$.

\section{Jaynes--Cummings Models in the Strong Coupling Regime}

In \cite{MFr} Frasca treated the Jaynes--Cummings model and developped some 
method to calculate Rabi frequencies in the strong coupling regime. 
We in this section generalize the model and method, and show 
that Rabi frequencies in our extended model are given by matrix elements 
of generalized coherent operators under the rotating--wave approximation. 
This gives a unified approach to them.

Let $\{\sigma_{1}, \sigma_{2}, \sigma_{3}\}$ be Pauli matrices and 
${\bf 1}_2$ a unit matrix : 
\begin{equation}
\sigma_{1} = 
\left(
  \begin{array}{cc}
    0& 1 \\
    1& 0
  \end{array}
\right), \quad 
\sigma_{2} = 
\left(
  \begin{array}{cc}
    0& -i \\
    i& 0
  \end{array}
\right), \quad 
\sigma_{3} = 
\left(
  \begin{array}{cc}
    1& 0 \\
    0& -1
  \end{array}
\right), \quad
{\bf 1}_2 = 
\left(
  \begin{array}{cc}
    1& 0 \\
    0& 1
  \end{array}
\right).
\end{equation}
The Hamiltonian adopted in \cite{MFr} is 
\begin{equation}
\mbox{(N)}\qquad H_{N}=\omega {\bf 1}_{2}\otimes a^{\dagger}a + 
\frac{\Delta}{2}\sigma_{3}\otimes {\bf 1} +
g\sigma_{1}\otimes (a^{\dagger}+a) 
\end{equation}
where $\omega$ is the frequency of the radiation mode, $\Delta$ 
the separation  between the two levels of the atom, $g$ the coupling 
between the radiation field and the atom. 

\par \noindent 
Moreover we want to treat the following Hamiltonians (our extension) 
\begin{eqnarray}
\mbox{(K)}\qquad H_{K}&=&\omega {\bf 1}_{2}\otimes K_{3} + 
\frac{\Delta}{2}\sigma_{3}\otimes {\bf 1}_{K} +
g\sigma_{1}\otimes (K_{+}+K_{-}), \\
\mbox{(J)}\qquad H_{J}&=&\omega {\bf 1}_{2}\otimes J_{3} + 
\frac{\Delta}{2}\sigma_{3}\otimes {\bf 1}_{J} +
g\sigma_{1}\otimes (J_{+}+J_{-}).
\end{eqnarray}

\par \noindent 
To treat these three cases at the same time we set 
\begin{equation}
\{L_{+},L_{-},L_{3}\}=
\left\{
\begin{array}{ll}
\mbox{(N)}\qquad \{a^{\dagger},a,N\} \\
\mbox{(K)}\qquad \{K_{+},K_{-},K_{3}\} \\
\mbox{(J)}\qquad \ \{J_{+},J_{-},J_{3}\} 
\end{array}
\right.
\end{equation}
and 
\begin{equation}
H=H_{0}+V
=\omega {\bf 1}_{2}\otimes L_{3} + 
\frac{\Delta}{2}\sigma_{3}\otimes {\bf 1}_{L} +
g\sigma_{1}\otimes (L_{+}+L_{-})
\end{equation}
where we have written $H$ instead of $H_{L}$ for simplicity. 

\par \noindent 
Mysteriously enough we cannot solve these simple models completely (maybe 
non--integrable), nevertheless we have found these models have a very rich 
structure. 

\par \noindent 
For these (non--integrable) models we usually have two perturbation 
approaches : 

\noindent{\bfseries Weak Coupling Regime} ($0 < g \ll \Delta$)\quad 
\begin{equation}
H_{0}=
\omega {\bf 1}_{2}\otimes L_{3} + 
\frac{\Delta}{2}\sigma_{3}\otimes {\bf 1}_{L}, \qquad 
V=g\sigma_{1}\otimes (L_{+}+L_{-}). 
\end{equation}

\noindent{\bfseries Strong Coupling Regime} ($0 < \Delta \ll g$)\quad 
\begin{equation}
H_{0}=
\omega {\bf 1}_{2}\otimes L_{3} + 
g\sigma_{1}\otimes (L_{+}+L_{-}), \qquad 
V=\frac{\Delta}{2}\sigma_{3}\otimes {\bf 1}_{L}. 
\end{equation}

\par \noindent 
In the following we consider only the strong coupling regime (see \cite{MSIII} 
for the weak one). 
First let us solve $H_{0}$ which is a relatively easy task. 

Let $W$ be a Walsh--Hadamard matrix 
\[
W=\frac{1}{\sqrt{2}}
\left(
  \begin{array}{cc}
    1& 1 \\
    1& -1
  \end{array}
\right)
=W^{-1}
\]
then we can diagonalize $\sigma_{1}$ by using this $H$ as 
$
\sigma_{1}=W\sigma_{3}W^{-1}
$.
The eigenvalues of $\sigma_{1}$ is $\{1,-1\}$ with eigenvectors 
\[
\ket{1}=\frac{1}{\sqrt{2}}
\left(
  \begin{array}{c}
    1 \\
    1
  \end{array}
\right), \quad 
\ket{-1}=\frac{1}{\sqrt{2}}
\left(
  \begin{array}{c}
    1 \\
    -1
  \end{array}
\right)
\quad \Longrightarrow \quad 
\ket{\lambda}=\frac{1}{\sqrt{2}}
\left(
  \begin{array}{c}
    1 \\
    \lambda
  \end{array}
\right).
\]
We note that
\begin{eqnarray}
&&\ket{1}\bra{1}=\frac{1}{2}
\left(
  \begin{array}{cc}
    1& 1 \\
    1& 1
  \end{array}
\right)=
W
\left(
  \begin{array}{cc}
    1& \\
     & 0
  \end{array}
\right)
W^{-1}, \nonumber \\
&&\ket{-1}\bra{-1}=\frac{1}{2}
\left(
  \begin{array}{cc}
    1& -1 \\
    -1& 1
  \end{array}
\right)=
W
\left(
  \begin{array}{cc}
    0& \\
     & 1
  \end{array}
\right)
W^{-1}, \nonumber \\
\Longrightarrow\quad 
&&\ket{\lambda}\bra{\lambda}=\frac{1}{2}
\left(
  \begin{array}{cc}
    1& \lambda \\
    \lambda& 1
  \end{array}
\right)=
W
\left(
  \begin{array}{cc}
    \frac{1+\lambda}{2}& \\
     & \frac{1-\lambda}{2}
  \end{array}
\right)
W^{-1}. \nonumber 
\end{eqnarray}

\par \noindent 
Then we have 
\begin{eqnarray}
\label{eq:Basic-Hamiltonian}
H_{0}&=&(W\otimes {\bf 1}_{L})
\left(
\omega {\bf 1}_{2}\otimes L_{3} + g\sigma_{3}\otimes (L_{+}+L_{-})
\right) (W^{-1}\otimes {\bf 1}_{L})  \nonumber \\
&=&(W\otimes {\bf 1}_{L})
\left(
  \begin{array}{cc}
    \omega L_{3} + g(L_{+}+L_{-})&  \\
    & \omega L_{3}-g(L_{+}+L_{-})
  \end{array}
\right)
(W^{-1}\otimes {\bf 1}_{L}). \nonumber \\
&=&\ket{1}\bra{1}\otimes \{\omega L_{3} + g(L_{+}+L_{-})\}
+ \ket{-1}\bra{-1}\otimes \{\omega L_{3} - g(L_{+}+L_{-})\}
    \nonumber \\
&=&\sum_{\lambda} \ket{\lambda}\bra{\lambda}\otimes 
\{\omega L_{3} + \lambda g(L_{+}+L_{-})\}  \nonumber \\
&=&\sum_{\lambda} \ket{\lambda}\bra{\lambda}\otimes 
\left\{
\mbox{e}^{-\frac{\lambda x}{2}(L_{+}-L_{-})}
\left(\Omega L_{3}\right)
\mbox{e}^{\frac{\lambda x}{2}(L_{+}-L_{-})}
\right\}    \nonumber \\
&=&\sum_{\lambda}
\left(
\ket{\lambda}\otimes \mbox{e}^{-\frac{\lambda x}{2}(L_{+}-L_{-})}
\right)
\left(\Omega L_{3}\right)
\left(
\bra{\lambda}\otimes \mbox{e}^{\frac{\lambda x}{2}(L_{+}-L_{-})}
\right) 
\end{eqnarray}
where we have used the following 

\noindent{\bfseries Key Formulas}\quad For $\lambda=\pm 1$ we have 
\begin{eqnarray}
\label{eq:N-formula}
&&(N)\quad 
\omega a^{\dagger}a + \lambda g(a^{\dagger}+a)
=\Omega \mbox{e}^{-\frac{\lambda x}{2}(a^{\dagger}-a)}
\left(N-\frac{g^2}{\omega^2}\right)
\mbox{e}^{\frac{\lambda x}{2}(a^{\dagger}-a)} \nonumber \\
&&\quad \qquad \quad  \mbox{where}\quad 
\Omega=\omega, \quad  x=2g/\omega, \\
\label{eq:K-formula}
&&(K)\quad 
\omega K_{3} + \lambda g(K_{+}+K_{-})
=\Omega\ \mbox{e}^{-\frac{\lambda x}{2}(K_{+}-K_{-})} K_{3}\ 
\mbox{e}^{\frac{\lambda x}{2}(K_{+}-K_{-})} \nonumber \\
&&\quad \qquad \quad  \mbox{where}\quad 
\Omega=\omega \sqrt{1-(2g/\omega)^{2}}, \quad 
x=\mbox{tanh}^{-1}(2g/\omega), \\
\label{eq:J-formula}
&&(J)\quad
\omega J_{3} + \lambda g(J_{+}+J_{-})
=\Omega\ \mbox{e}^{-\frac{\lambda x}{2}(J_{+}-J_{-})} J_{3}\ 
\mbox{e}^{\frac{\lambda x}{2}(J_{+}-J_{-})} \nonumber \\
&&\quad \qquad \quad  \mbox{where}\quad 
\Omega=\omega \sqrt{1+(2g/\omega)^{2}}, \quad 
x=\mbox{tan}^{-1}(2g/\omega). 
\end{eqnarray}

\par \noindent 
The proof is not difficult, so we leave it to the readers. 
That is, we could diagonalize the Hamiltonian $H_{0}$. This is two--fold 
degenerate and its eigenvalues and eigenvectors are given respectively 
\begin{equation}
\label{eq:Eigenvalues-Eigenvectors}
(\mbox{Eigenvalues},\mbox{Eigenvectors})=
\left\{
\begin{array}{ll}
(N)\quad \omega n - \frac{g^2}{\omega},\quad \quad 
\ket{\lambda}\otimes 
\mbox{e}^{-\frac{\lambda x}{2}(a^{\dagger}-a)}\ket{n} \\
(K)\quad \Omega (K+n),\quad 
\ket{\lambda}\otimes 
\mbox{e}^{-\frac{\lambda x}{2}(K_{+}-K_{-})}\ket{K,n} \\
(J)\quad \Omega (-J+n),\quad 
\ket{\lambda}\otimes 
\mbox{e}^{-\frac{\lambda x}{2}(J_{+}-J_{-})}\ket{J,n} \\
\end{array}
\right.
\end{equation}
for $\lambda=\pm 1$ and $n \in \futon \cup \{0\}$. 
For the latter convenience we set 
\begin{equation}
\label{eq:eigenvalues-vectors}
\mbox{Eigenvalues}=\{E_{n}\}, \quad 
\mbox{Eigenvectors}=\{\ket{\{\lambda, n\}}\}. 
\end{equation}
Then (\ref{eq:Basic-Hamiltonian}) can be written as 
\begin{equation}
\label{eq:Basic-Hamiltonian-2}
H_{0}
=\sum_{\lambda}\sum_{n}E_{n} \ket{\{\lambda, n\}}\bra{\{\lambda, n\}}.
\end{equation}

\par \vspace{5mm} 
Next we would like to solve the following Schr{\"o}dinger equation : 
\begin{equation}
\label{eq:full-equation}
i\frac{d}{dt}\Psi=H\Psi=\left(H_{0}+\frac{\Delta}{2}
\sigma_{3}\otimes {\bf 1}_{L}\right)\Psi, 
\end{equation}
where we have set $\hbar=1$ for simplicity. 
To solve this equation we appeal to the method of constant variation. 
First let us solve 
\begin{equation}
\label{eq:partial-equation}
i\frac{d}{dt}\Psi=H_{0}\Psi, 
\end{equation}
which general solution is given by 
\begin{equation}
\label{eq:partial-solution}
  \Psi(t)=U_{0}(t)\Psi_{0}=\mbox{e}^{-itH_{0}}\Psi_{0}
\end{equation}
where $\Psi_{0}$ is a constant state. It is easy to see from 
(\ref{eq:Basic-Hamiltonian-2})
\begin{equation}
\label{eq:Basic-Unitary}
U_{0}(t)=\mbox{e}^{-itH_{0}}
=\sum_{\lambda}\sum_{n}\mbox{e}^{-itE_{n}}
\ket{\{\lambda, n\}}\bra{\{\lambda, n\}}.
\end{equation}

The method of constant variation goes as follows. Changing like \quad 
$
\Psi_{0} \longrightarrow \Psi_{0}(t),
$ \quad 
we insert (\ref{eq:partial-solution}) into (\ref{eq:full-equation}). 
After some algebra we obtain 
\begin{equation}
\label{eq:sub-equation}
i\frac{d}{dt}\Psi_{0}
=\frac{\Delta}{2}{U_0}^{\dagger}(\sigma_{3}\otimes {\bf 1}_{L}){U_0}\Psi_{0}.
\end{equation}
We have only to solve this equation. If we set 
\begin{equation}
\label{eq:Frasca-hamiltonian}
H_{F}=\frac{\Delta}{2}{U_0}^{\dagger}(\sigma_{3}\otimes {\bf 1}_{L}){U_0}, 
\end{equation}
then we have easily from (\ref{eq:Basic-Unitary})
\begin{eqnarray}
H_{F}&=&\frac{\Delta}{2}
\sum_{\lambda,\mu}\sum_{m, n}
\mbox{e}^{it(E_m-E_n)} 
\bra{\{\lambda,m\}}(\sigma_{3}\otimes {\bf 1}_{L})\ket{\{\mu,n\}} \ 
\ket{\{\lambda, m\}}\bra{\{\mu, n\}} \nonumber \\
&=&\frac{\Delta}{2}
\sum_{\lambda}\sum_{m, n}
\mbox{e}^{it\Omega(m-n)} 
\braa{m}\mbox{e}^{{\lambda x}(L_{+}-L_{-})}\kett{n} \ 
\ket{\{\lambda, m\}}\bra{\{-\lambda, n\}} 
\end{eqnarray}
where we have used the relation $\bra{\lambda}\sigma_{3}=\bra{-\lambda}$. \ 
Remind that $\kett{n}$ is respectively 
\[
\kett{n}=
\left\{
\begin{array}{ll}
(N)\qquad \ket{n} \\
(K)\qquad \ket{K,n} \\
(J)\qquad \ket{J,n}
\end{array}
\right.
\]

\par \noindent 
In this stage we meet {\bf matrix elements of the coherent and generalized 
coherent operators} $\mbox{e}^{{\lambda x}(L_{+}-L_{-})}$ 
in section 3 ($z={\bar z}={\lambda x}$).

\par \noindent
Here we divide $H_{F}$ into two parts
\[
H_{F}={H_{F}}^{'}+{H_{F}}^{''}
\]
where 
\begin{eqnarray}
\label{eq:SecondHamiltonian-1}
{H_{F}}^{'}&=&\frac{\Delta}{2}\sum_{\lambda}\sum_{n}
\braa{n}\mbox{e}^{{\lambda x}(L_{+}-L_{-})}\kett{n} \ 
\ket{\{\lambda, n\}}\bra{\{-\lambda, n\}}, \\
\label{eq:SecondHamiltonian-2}
{H_{F}}^{''}&=&\frac{\Delta}{2}\sum_{\lambda}
\sum_{\stackrel{\scriptstyle m,n}{m\ne n}}
\mbox{e}^{it\Omega(m-n)} 
\braa{m}\mbox{e}^{{\lambda x}(L_{+}-L_{-})}\kett{n} \ 
\ket{\{\lambda, m\}}\bra{\{-\lambda, n\}}.
\end{eqnarray}
Noting 
\[
\braa{n}\mbox{e}^{x(L_{+}-L_{-})}\kett{n}=
\braa{n}\mbox{e}^{-x(L_{+}-L_{-})}\kett{n}
\]
by the results in section 3, ${H_{F}}^{'}$ can be written as
\[
{H_{F}}^{'}=\frac{\Delta}{2}\sum_{n}
\braa{n}\mbox{e}^{x(L_{+}-L_{-})}\kett{n} 
\left\{ 
\ket{\{1, n\}}\bra{\{-1, n\}} + \ket{\{-1, n\}}\bra{\{1, n\}}
\right\}, 
\]
from which we can diagonalize ${H_{F}}^{'}$ as
\begin{equation}
\label{eq:diagonal-hamiltonian}
{H_{F}}^{'}=\frac{\Delta}{2}\sum_{n}\sum_{\sigma}
\braa{n}\mbox{e}^{x(L_{+}-L_{-})}\kett{n}\sigma \ 
\ket{\{\sigma,{\psi}_{n}\}} \bra{\{\sigma,{\psi}_{n}\}} 
\end{equation}
if we define a new basis 
\[
|\{\sigma,{\psi}_{n}\}\rangle = \frac{1}{\sqrt{2}}
(\sigma \ket{\{1, n\}} + \ket{\{-1, n\}}), 
\qquad \sigma=\pm 1. 
\]
These states can be seen as so--called Schr{\"o}dinger cat states, 
\cite{WPS}. From these we have 
\begin{eqnarray}
\ket{\{1, n\}}&=&\frac{1}{\sqrt{2}}
\{|\{1,{\psi}_{n}\}\rangle - |\{-1,{\psi}_{n}\}\rangle \}, \nonumber \\
\ket{\{-1, n\}}&=&\frac{1}{\sqrt{2}}
\{|\{1,{\psi}_{n}\}\rangle + |\{-1,{\psi}_{n}\}\rangle \}.  \nonumber 
\end{eqnarray}
Inserting these equations into (\ref{eq:SecondHamiltonian-2}) and 
taking some algebras we obtain  
\begin{eqnarray}
\label{eq:non-diagonal-hamiltonian}
{H_{F}}^{''}&=&\frac{\Delta}{2}
\sum_{\stackrel{\scriptstyle m,n}{m\ne n}}\sum_{\sigma,\sigma^{'}}
\mbox{e}^{it\Omega(m-n)} 
\left\{
\braa{m}\mbox{e}^{x(L_{+}-L_{-})}\kett{n}
\frac{\sigma}{2}\ 
|\{\sigma, {\psi}_{m}\}\rangle \langle\{\sigma^{'}, {\psi}_{n}\}|\ + 
\right.          \nonumber \\
&&\left. \qquad \qquad \qquad \qquad \quad 
\braa{m}\mbox{e}^{-x(L_{+}-L_{-})}\kett{n}
\frac{\sigma^{'}}{2}\ 
|\{\sigma, {\psi}_{m}\}\rangle \langle\{\sigma^{'}, {\psi}_{n}\}|
\right\}. 
\end{eqnarray}
For simplicity in (\ref{eq:diagonal-hamiltonian})
we set in the following 
\begin{equation}
E_{n,\sigma}=\frac{\Delta}{2}\sigma 
\braa{n}\mbox{e}^{x(L_{+}-L_{-})}\kett{n}, 
\end{equation}
then 
\begin{equation}
E_{n,\sigma}=
\left\{
\begin{array}{ll}
(N)\quad \frac{\Delta}{2}\sigma 
\mbox{e}^{-\frac{2g^2}{\omega^2}}L_{n}\left(\frac{4g^2}{\omega^2}\right) \\
(K)\quad \frac{\Delta}{2}\sigma 
\frac{n!}{(2K)_{n}}(1+|\kappa|^2)^{-K-n}
F_{n}(|\kappa|^2:2K)\quad \mbox{where}\quad \kappa=\mbox{sinh}(x)  \\
(J)\quad \frac{\Delta}{2}\sigma 
\frac{n!}{{}_{2J}P_n}(1-|\kappa|^2)^{J-n}
F_{n}(|\kappa|^2:2J)\quad \mbox{where}\quad \kappa=\mbox{sin}(x)  
\end{array}
\right.
\end{equation}
from (\ref{eq:Eigenvalues-Eigenvectors}) and the results in sectin 3.1. 
Now let us solve (\ref{eq:sub-equation})
\[
i\frac{d}{dt}\Psi_{0}
=\frac{\Delta}{2} H_{F}\Psi_{0}
=\frac{\Delta}{2} ({H_{F}}^{'}+{H_{F}}^{''})\Psi_{0}.
\]
For that if we set $\Psi_{0}(t)$ as 
\begin{equation}
\label{eq:full-ansatz}
\Psi_{0}(t)=\sum_{\sigma}\sum_{n}\mbox{e}^{-itE_{n,\sigma}}a_{n,\sigma}(t)
|\{\sigma,{\psi}_{n}\}\rangle ,
\end{equation}
then we have a set of complicated equations with respect to 
$\{a_{n,\sigma}\}$, see \cite{MFr}. 
But it is almost impossible to solve them. Therefore we make 
a daring assumption : for $m < n$ 
\begin{equation}
\label{eq:daring-ansatz}
\Psi_{0}(t)=
\sum_{\sigma}\mbox{e}^{-itE_{m,\sigma}}a_{m,\sigma}(t)
|\{\sigma,{\psi}_{m}\}\rangle 
+ 
\sum_{\sigma}\mbox{e}^{-itE_{n,\sigma}}a_{n,\sigma}(t)
|\{\sigma,{\psi}_{n}\}\rangle.
\end{equation}
That is, we consider only two terms with respect to $\{n | n \geq 0\}$. 
After some algebras we obtain 
\begin{eqnarray}
\label{eq:m,n-equations}
i\frac{d}{dt}a_{m,\sigma}&=&
\frac{\Delta}{2}\sum_{\sigma^{'}}
\mbox{e}^{-it(E_{n,\sigma^{'}}-E_{m,\sigma})}
\mbox{e}^{it\Omega(m-n)}
\left\{
\braa{m}\mbox{e}^{x(L_{+}-L_{-})}\kett{n}\frac{\sigma}{2}
+ 
\braa{m}\mbox{e}^{-x(L_{+}-L_{-})}\kett{n}\frac{\sigma^{'}}{2}
\right\}a_{n,\sigma^{'}},  \nonumber \\
&&{} \\
i\frac{d}{dt}a_{n,\sigma}&=&
\frac{\Delta}{2}\sum_{\sigma^{'}}
\mbox{e}^{-it(E_{m,\sigma^{'}}-E_{n,\sigma})}
\mbox{e}^{it\Omega(n-m)}
\left\{
\braa{n}\mbox{e}^{x(L_{+}-L_{-})}\kett{m}\frac{\sigma}{2}
+ 
\braa{n}\mbox{e}^{-x(L_{+}-L_{-})}\kett{m}\frac{\sigma^{'}}{2}
\right\}a_{m,\sigma^{'}}. \nonumber 
\end{eqnarray}
But we cannot still solve the above equations exactly (see Appendix), 
so let us make so--called rotating--wave approximation.  
The resonance condition is 
\begin{equation}
-(E_{n,\sigma^{'}}-E_{m,\sigma})+(m-n)\Omega=0\quad \Longrightarrow \quad 
E_{n,\sigma^{'}}-E_{m,\sigma}=(m-n)\Omega 
\end{equation}
for some $\sigma$ and $\sigma^{'}$, and we reject the remaining term in 
(\ref{eq:m,n-equations}).  Then we obtain simple equations : 

\noindent{\bfseries Interband Transition Case} ($\sigma \ne \sigma^{'}$)\qquad 
$E_{n,-\sigma}-E_{m,\sigma}=(m-n)\Omega$ 
\begin{eqnarray}
\label{eq:interband-m,n-equations}
i\frac{d}{dt}a_{m,\sigma}&=&
\frac{\Delta}{2}
\left\{
\braa{m}\mbox{e}^{x(L_{+}-L_{-})}\kett{n}\frac{\sigma}{2}
-
\braa{m}\mbox{e}^{-x(L_{+}-L_{-})}\kett{n}\frac{\sigma}{2}
\right\}a_{n,-\sigma} \nonumber \\
&=&
\frac{\Delta}{2}\sigma
\braa{m}\mbox{sinh}\left(x(L_{+}-L_{-})\right)\kett{n} 
a_{n,-\sigma},  \nonumber \\
&&{} \\
i\frac{d}{dt}a_{n,-\sigma}&=&
\frac{\Delta}{2}\sum_{\sigma^{'}}
\left\{-
\braa{n}\mbox{e}^{x(L_{+}-L_{-})}\kett{m}\frac{\sigma}{2}
+
\braa{n}\mbox{e}^{-x(a^{\dagger}-a)}\kett{m}\frac{\sigma}{2}
\right\}a_{m,\sigma} \nonumber \\
&=&
-\frac{\Delta}{2}\sigma
\braa{n}\mbox{sinh}\left(x(L_{+}-L_{-})\right)\kett{m} 
a_{m,\sigma}.  \nonumber 
\end{eqnarray}
\noindent{\bfseries Intraband Transition Case} ($\sigma = \sigma^{'}$)\qquad 
$E_{n,\sigma}-E_{m,\sigma}=(m-n)\Omega$ 
\begin{eqnarray}
\label{eq:intraband-m,n-equations}
i\frac{d}{dt}a_{m,\sigma}&=&
\frac{\Delta}{2}
\left\{
\braa{m}\mbox{e}^{x(L_{+}-L_{-})}\kett{n}\frac{\sigma}{2}
+
\braa{m}\mbox{e}^{-x(L_{+}-L_{-})}\kett{n}\frac{\sigma}{2}
\right\}a_{n,\sigma} \nonumber \\
&=&
\frac{\Delta}{2}\sigma
\braa{m}\mbox{cosh}\left(x(L_{+}-L_{-})\right)\kett{n} 
a_{n,\sigma},  \nonumber \\
&&{} \\
i\frac{d}{dt}a_{n,\sigma}&=&
\frac{\Delta}{2}\sum_{\sigma^{'}}
\left\{
\braa{n}\mbox{e}^{x(L_{+}-L_{-})}\kett{m}\frac{\sigma}{2}
+
\braa{n}\mbox{e}^{-x(L_{+}-L_{-})}\kett{m}\frac{\sigma}{2}
\right\}a_{m,\sigma} \nonumber \\
&=&
\frac{\Delta}{2}\sigma
\braa{n}\mbox{cosh}\left(x(L_{+}-L_{-})\right)\kett{m} 
a_{m,\sigma}.  \nonumber 
\end{eqnarray}

For simplicity we set
\begin{equation}
{\cal R}=\Delta
\braa{n}\mbox{sinh}\left(x(L_{+}-L_{-})\right)\kett{m}, 
\quad 
{\cal R}^{'}=\Delta
\braa{n}\mbox{cosh}\left(x(L_{+}-L_{-})\right)\kett{m}, 
\end{equation}
then 
\[
\Delta
\braa{m}\mbox{sinh}\left(x(L_{+}-L_{-})\right)\kett{n} 
=-{\cal R}, \quad 
\Delta
\braa{m}\mbox{cosh}\left(x(L_{+}-L_{-})\right)\kett{n}
={\cal R}^{'}. 
\]
These are two Rabi frequencies as shown in the following. 
It is important that Rabi frequencies in our models are given by 
{\bf matrix elements of coherent and generalized coherent operators} !  

\par \noindent 
By making use of the results in section 3 and 
(\ref{eq:N-formula}), (\ref{eq:K-formula}), (\ref{eq:J-formula}) we have 
\begin{eqnarray}
&&(N)\  
\left\{
\begin{array}{ll}
{\cal R}=
\frac{\Delta}{2}\sqrt{\frac{m!}{n!}}\left(\frac{2g}{\omega}\right)^{n-m}
\mbox{e}^{-\frac{2g^2}{\omega^2}}{L_{m}}^{(n-m)}
\left(\frac{4g^2}{\omega^2}\right)\{1-(-1)^{n-m}\} \\
{\cal R}^{'}=
\frac{\Delta}{2}\sqrt{\frac{m!}{n!}}\left(\frac{2g}{\omega}\right)^{n-m}
\mbox{e}^{-\frac{2g^2}{\omega^2}}{L_{m}}^{(n-m)}
\left(\frac{4g^2}{\omega^2}\right)\{1+(-1)^{n-m}\}
\end{array}
\right.           \\
&&{} \nonumber \\
&&(K)\ 
\left\{
\begin{array}{ll}
{\cal R}=
\frac{\Delta}{2}
\sqrt{\frac{n!m!}{(2K)_n(2K)_m}}
{\kappa^{n-m}}(1+\kappazetta^2)^{-K-\frac{n+m}{2}}
{F_{m}}^{(n-m)}(\kappazetta^2:2K)\{1-(-1)^{n-m}\} \\
{\cal R}^{'}=
\frac{\Delta}{2}
\sqrt{\frac{n!m!}{(2K)_n(2K)_m}}
{\kappa^{n-m}}(1+\kappazetta^2)^{-K-\frac{n+m}{2}}
{F_{m}}^{(n-m)}(\kappazetta^2:2K)\{1+(-1)^{n-m}\}
\end{array}
\right.           \\
&&\qquad \qquad \qquad \qquad \qquad \mbox{where}\quad 
\kappa=\mbox{sinh}\left(x\right)\quad \mbox{with}\quad 
x=\mbox{tanh}^{-1}\left(\frac{2g}{\omega}\right)  
\nonumber \\
&&(J)\  
\left\{
\begin{array}{ll}
{\cal R}=
\frac{\Delta}{2}
\sqrt{\frac{n!m!}{{}_{2J}P_n {}_{2J}P_m}}
{\kappa^{n-m}}(1-\kappazetta^2)^{J-\frac{n+m}{2}}
{F_{m}}^{(n-m)}(\kappazetta^2:2J)\{1-(-1)^{n-m}\} \\
{\cal R}^{'}=
\frac{\Delta}{2}
\sqrt{\frac{n!m!}{{}_{2J}P_n {}_{2J}P_m}}
{\kappa^{n-m}}(1-\kappazetta^2)^{J-\frac{n+m}{2}}
{F_{m}}^{(n-m)}(\kappazetta^2:2J)\{1+(-1)^{n-m}\}
\end{array}
\right.  \\
&&\qquad \qquad \qquad \qquad \qquad  \mbox{where}\quad 
\kappa=\mbox{sin}\left(x\right)\quad \mbox{with}\quad  
x=\mbox{tan}^{-1}\left(\frac{2g}{\omega}\right)  
\nonumber 
\end{eqnarray}
From these we find a constraint between $m$ and $n$ : 

\noindent{\bfseries Interband Case}\quad 
$n-m=2N-1$\quad  $\Longrightarrow$ \quad $n=m+2N-1$ \quad 
for \quad $N \in \futon$, 

\noindent{\bfseries Intraband Case}\quad 
$n-m=2N$\quad $\Longrightarrow$ \quad $n=m+2N$ \quad for \quad 
$N \in \futon$.

Now let us solve (\ref{eq:interband-m,n-equations}) and 
(\ref{eq:intraband-m,n-equations}). 
\begin{eqnarray}
i\frac{d}{dt}
\left(
\begin{array}{c}
a_{m,\sigma} \\
a_{n,-\sigma}
\end{array}
\right)
&=&
\left(
\begin{array}{cc}
0& -\sigma \frac{{\cal R}}{2} \\
-\sigma \frac{{\cal R}}{2}& 0
\end{array}
\right)
\left(
\begin{array}{c}
a_{m,\sigma} \\
a_{n,-\sigma}
\end{array}
\right),  \nonumber \\
&&{} \\
i\frac{d}{dt}
\left(
\begin{array}{c}
a_{m,\sigma} \\
a_{n,\sigma}
\end{array}
\right)
&=&
\left(
\begin{array}{cc}
0& \sigma \frac{{\cal R}^{'}}{2} \\
\sigma \frac{{\cal R}^{'}}{2}& 0
\end{array}
\right)
\left(
\begin{array}{c}
a_{m,\sigma} \\
a_{n,\sigma}
\end{array}
\right),    \nonumber 
\end{eqnarray}
so their solutions are given by 
\begin{eqnarray}
\left(
\begin{array}{c}
a_{m,\sigma}(t) \\
a_{n,-\sigma}(t)
\end{array}
\right)
&=&
\left(
\begin{array}{cc}
\mbox{cos}(\frac{{\cal R}}{2}t)& i\sigma\mbox{sin}(\frac{{\cal R}}{2}t)
\\
i\sigma\mbox{sin}(\frac{{\cal R}}{2}t)& \mbox{cos}(\frac{{\cal R}}{2}t)
\end{array}
\right)
\left(
\begin{array}{c}
a_{m,\sigma}(0) \\
a_{n,-\sigma}(0)
\end{array}
\right),  \nonumber \\
&&{} \\
\left(
\begin{array}{c}
a_{m,\sigma}(t) \\
a_{n,\sigma}(t)
\end{array}
\right)
&=&
\left(
\begin{array}{cc}
\mbox{cos}(\frac{{\cal R}^{'}}{2}t)&
-i\sigma\mbox{sin}(\frac{{\cal R}^{'}}{2}t) \\
-i\sigma\mbox{sin}(\frac{{\cal R}^{'}}{2}t)& 
\mbox{cos}(\frac{{\cal R}^{'}}{2}t)
\end{array}
\right)
\left(
\begin{array}{c}
a_{m,\sigma}(0) \\
a_{n,\sigma}(0)
\end{array}
\right).  \nonumber 
\end{eqnarray}

\par \noindent 
We have obtained some solutions under the rotating--wave approximation. 
Now it may be suited to compare our results with a recent experimental  
finding in \cite{NPT}, but this is beyond our scope. See \cite{MFr}.

Let us conclude this section by a comment. 
Our ansatz (\ref{eq:daring-ansatz}) to solve the equation is too 
restrictive. We want to use (\ref{eq:full-ansatz}) to solve the 
equation, but it is very hard at this stage. 

\noindent{\bfseries Problem}\quad Find more dynamic methods ! 

\section{Quantum Computation}

Let us reconsider the results in the preceding section in the light of 
Quantum Computation. Remind once more that the following arguments are based 
on the {\bf rotating--wave approximation}. 

\vspace{5mm}
\noindent{\bfseries Interband Case}\ $\bf{(\sigma=1)}$

\begin{eqnarray}
i\frac{d}{dt}
\left(
\begin{array}{c}
a_{m,1} \\
a_{m,-1} \\
a_{n,1}\\
a_{n,-1}
\end{array}
\right)
&=&
\left(
\begin{array}{cccc}
0& 0& 0& -\frac{{\cal R}}{2} \\
0& 0& 0& 0 \\
0& 0& 0& 0 \\
-\frac{{\cal R}}{2}& 0& 0& 0
\end{array}
\right)
\left(
\begin{array}{c}
a_{m,1} \\
a_{m,-1} \\
a_{n,1}\\
a_{n,-1}
\end{array}
\right). 
\end{eqnarray}
The solution is 
\begin{eqnarray}
\label{eq:solution-m-1}
\left(
\begin{array}{c}
a_{m,1}(t) \\
a_{m,-1}(t) \\
a_{n,1}(t) \\
a_{n,-1}(t)
\end{array}
\right)
&=&
\left(
\begin{array}{cccc}
\mbox{cos}(\frac{{\cal R}}{2}t)& 0 &0 &
i\mbox{sin}(\frac{{\cal R}}{2}t)
\\
0& 1& 0& 0 \\
0& 0& 1& 0 \\
i\mbox{sin}(\frac{{\cal R}}{2}t)& 0& 0&
\mbox{cos}(\frac{{\cal R}}{2}t)
\end{array}
\right)
\left(
\begin{array}{c}
a_{m,1}(0) \\
a_{m,-1}(0) \\
a_{n,1}(0) \\
a_{n,-1}(0)
\end{array}
\right).
\end{eqnarray}

\vspace{3mm}
\noindent{\bfseries Interband Case}\ $\bf{(\sigma=-1)}$

\begin{eqnarray}
i\frac{d}{dt}
\left(
\begin{array}{c}
a_{m,1} \\
a_{m,-1} \\
a_{n,1}\\
a_{n,-1}
\end{array}
\right)
&=&
\left(
\begin{array}{cccc}
0& 0& 0& 0 \\
0& 0& \frac{{\cal R}}{2}& 0 \\
0& \frac{{\cal R}}{2}& 0& 0 \\
0& 0& 0& 0
\end{array}
\right)
\left(
\begin{array}{c}
a_{m,1} \\
a_{m,-1} \\
a_{n,1}\\
a_{n,-1}
\end{array}
\right). 
\end{eqnarray}
The solution is 
\begin{eqnarray}
\label{eq:solution-m-(-1)}
\left(
\begin{array}{c}
a_{m,1}(t) \\
a_{m,-1}(t) \\
a_{n,1}(t) \\
a_{n,-1}(t)
\end{array}
\right)
&=&
\left(
\begin{array}{cccc}
1& 0& 0& 0 \\
0& \mbox{cos}(\frac{{\cal R}}{2}t)& 
-i\mbox{sin}(\frac{{\cal R}}{2}t)& 0 \\
0& -i\mbox{sin}(\frac{{\cal R}}{2}t)& 
\mbox{cos}(\frac{{\cal R}}{2}t)& 0 \\ 
0& 0& 0& 1
\end{array}
\right)
\left(
\begin{array}{c}
a_{m,1}(0) \\
a_{m,-1}(0) \\
a_{n,1}(0) \\
a_{n,-1}(0)
\end{array}
\right).
\end{eqnarray}

\vspace{5mm}
\noindent{\bfseries Intraband Case}\ $\bf{(\sigma=1)}$

\begin{eqnarray}
i\frac{d}{dt}
\left(
\begin{array}{c}
a_{m,1} \\
a_{m,-1} \\
a_{n,1}\\
a_{n,-1}
\end{array}
\right)
&=&
\left(
\begin{array}{cccc}
0& 0& \frac{{\cal R}^{'}}{2}& 0 \\
0& 0& 0& 0 \\
\frac{{\cal R}^{'}}{2}& 0& 0& 0 \\
0& 0& 0& 0 
\end{array}
\right)
\left(
\begin{array}{c}
a_{m,1} \\
a_{m,-1} \\
a_{n,1}\\
a_{n,-1}
\end{array}
\right). 
\end{eqnarray}
The solution is 
\begin{eqnarray}
\label{eq:solution-n-1}
\left(
\begin{array}{c}
a_{m,1}(t) \\
a_{m,-1}(t) \\
a_{n,1}(t) \\
a_{n,-1}(t)
\end{array}
\right)
&=&
\left(
\begin{array}{cccc}
\mbox{cos}(\frac{{\cal R}^{'}}{2}t)& 0& 
-i\mbox{sin}(\frac{{\cal R}^{'}}{2}t)& 0 \\
0& 1& 0& 0 \\
-i\mbox{sin}(\frac{{\cal R}^{'}}{2}t)& 0& 
\mbox{cos}(\frac{{\cal R}^{'}}{2}t)& 0 \\
0& 0& 0& 1 
\end{array}
\right)
\left(
\begin{array}{c}
a_{m,1}(0) \\
a_{m,-1}(0) \\
a_{n,1}(0) \\
a_{n,-1}(0)
\end{array}
\right).
\end{eqnarray}

\vspace{3mm}
\noindent{\bfseries Intraband Case}\ $\bf{(\sigma=-1)}$

\begin{eqnarray}
i\frac{d}{dt}
\left(
\begin{array}{c}
a_{m,1} \\
a_{m,-1} \\
a_{n,1}\\
a_{n,-1}
\end{array}
\right)
&=&
\left(
\begin{array}{cccc}
0& 0& 0& 0 \\
0& 0& 0& -\frac{{\cal R}^{'}}{2} \\
0& 0& 0& 0 \\
0& -\frac{{\cal R}^{'}}{2}& 0& 0 \\ 
\end{array}
\right)
\left(
\begin{array}{c}
a_{m,1} \\
a_{m,-1} \\
a_{n,1}\\
a_{n,-1}
\end{array}
\right). 
\end{eqnarray}
The solution is 
\begin{eqnarray}
\label{eq:solution-n-(-1)}
\left(
\begin{array}{c}
a_{m,1}(t) \\
a_{m,-1}(t) \\
a_{n,1}(t) \\
a_{n,-1}(t)
\end{array}
\right)
&=&
\left(
\begin{array}{cccc}
1& 0& 0& 0 \\
0& \mbox{cos}(\frac{{\cal R}^{'}}{2}t)& 0&
i\mbox{sin}(\frac{{\cal R}^{'}}{2}t) \\
0& 0& 1& 0 \\
0& i\mbox{sin}(\frac{{\cal R}^{'}}{2}t)& 0&
\mbox{cos}(\frac{{\cal R}^{'}}{2}t)
\end{array}
\right)
\left(
\begin{array}{c}
a_{m,1}(0) \\
a_{m,-1}(0) \\
a_{n,1}(0) \\
a_{n,-1}(0)
\end{array}
\right).
\end{eqnarray}

If we can identify (\ref{eq:daring-ansatz}) with an element in two--qubit 
space 
\begin{equation}
a_{m,1}(t)|00\rangle + a_{m,-1}(t)|01\rangle + 
a_{n,1}(t)|10\rangle + a_{n,-1}(t)|11\rangle \ \in \ 
\fukuso^{2}\otimes \fukuso^{2}
\end{equation}
where $\fukuso^{2}=\mbox{Vect}_{\fukuso}\{\ket{0}, \ket{1}\}$, 
then the solutions (\ref{eq:solution-m-1}), (\ref{eq:solution-m-(-1)}), 
(\ref{eq:solution-n-1}), (\ref{eq:solution-n-(-1)}) are kinds of 
controlled unitary operations (gates) which play a crucial role in 
Quantum Computation, see for example \cite{KF8}.
For example, (\ref{eq:solution-n-(-1)}) is just one of controlled unitary 
gates expressed graphically as 
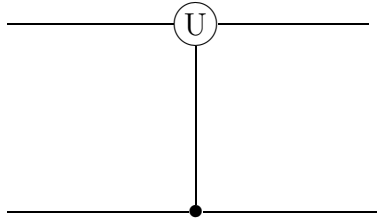
\begin{center}
\setlength{\unitlength}{1mm}  
\begin{picture}(120,50)
\put(25,35){\line(1,0){22}}   
\put(53,35){\line(1,0){20}}   
\put(25,10){\line(1,0){24}}   
\put(51,10){\line(1,0){23}}   
\put(50,11){\line(0,1){21}}   
\put(47,5){\makebox(6,10){$\bullet$}} 
\put(50,35){\circle{6}}               
\put(47,30){\makebox(6,10){U}}         
\end{picture}
\end{center}
We note here that controlled unitary gates above are written down as 
\begin{equation}
\mbox{C--Unitary}=
\left(
\begin{array}{cccc}
1& 0& 0& 0 \\
0& u_{11}& 0& u_{12} \\
0& 0& 1& 0 \\
0& u_{21}& 0& u_{22}
\end{array}
\right)
\end{equation}
for 
\[
U=
\left(
\begin{array}{cc}
u_{11}& u_{12} \\
u_{21}& u_{22}
\end{array}
\right)\ \in \ U(2) .
\]

\par \vspace{5mm} \noindent 
A comment is in order.\ \ 
\cite{several-1} (and \cite{several-2}) is considering the same subject. 
However the authors in them treated it in the weak coupling regime, 
while we treated it in the strong coupling regime. It is very interesting 
to investigate a deep relation (connection) between them.

\section{Discussion}

One of motivations of this study is to apply our results to 
Holonomic Quantum Computation developped by Italian group (Pachos, 
Rasetti and Zanardi) and the author, see \cite{ZR}, \cite{PZR}, \cite{PC}, 
\cite{PZ} and \cite{KF0}---\cite{KF4} and recent \cite{DL}, \cite{DL2}. 

In this theory we usually use the effective Hamiltonian of 
a single--mode field of Kerr medium 
\begin{equation}
\label{eq:effective-Hamiltonian}
H_{0}=XN(N-1), \quad N=a^{\dagger}a \quad \mbox{where}\quad X\quad 
\mbox{is a constant}
\end{equation}
as a background and the real Hamiltonian is in one--qubit case given by 
\begin{equation}
\label{eq:real-Hamiltonian}
H(z,w)=W(z,w)H_{0}W^{-1}(z,w)
\end{equation}
where $W$ is a product of coherent operator $U(z)$ and squeezed one $S(w)$ 
in section 2. In the above Hamiltonian $H_{0}$ the 
zero--eigenvalue is two--fold degenerate whose eigenvectors are 
$\ket{0}$ and $\ket{1}$. We set $\ket{vac}=(\ket{0},\ket{1})$. Then we can 
construct a connection form $\cala$ on the parameter space $\{(z,w)\in 
\fukuso^{2}\}$  as 
\begin{equation}
\label{eq:connection-form}
\cala=\bra{vac}W^{-1}dW\ket{vac}
\end{equation}
from (\ref{eq:real-Hamiltonian})\  where 
$d=dz\frac{\partial}{\partial z}+dw\frac{\partial}{\partial w}$. 
By making use of this connection we can construct a holonomy group 
$Hol(\cala)$ ($ \subseteq U(2)$) which is in this case equal to $U(2)$. 
In Holonomic Quantum Computation we use this holonomy group as unitary 
operations in Quantum Computation. 
The point at issue is that we use not full property of the Hamiltonian 
but only property of the zero--eigenvalue. 

\par \vspace{2mm}
By the way, the Hamiltonian $H_{F}$ in (\ref{eq:Frasca-hamiltonian}) 
\[
H_{F}=\frac{\Delta}{2}
{U_0}^{-1}(\sigma_{3}\otimes {\bf 1}_{L}){U_0}\quad 
\mbox{where}\quad  L=(N)\ \ \mbox{or}\ \ (K)\ \ \mbox{or}\ \ (J) 
\]
is very similar to (\ref{eq:real-Hamiltonian}). This system is always 
two--fold degenerate. Then a natural question arises :

\begin{flushleft}
{\bf Problem}\quad Is it possible to perform a holonomic quantum 
computation by combining the systems \{(N), (K), (J)\} ?
\end{flushleft}
This is a very interesting and challenging problem.

\par \vspace{5mm}
\noindent{\em Acknowledgment.}\\
The author wishes to thank Marco Frasca for his helpful comments and 
suggestions.

\par \vspace{5mm}
\begin{center}
 \begin{Large}
   {\bf Appendix}
 \end{Large}
\end{center}
\begin{flushleft}
\begin{Large}
{\bf On Equations (\ref{eq:m,n-equations})}
\end{Large}
\end{flushleft}
Here let us write down full equations of (\ref{eq:m,n-equations}) with 
matrix equation form : 
\begin{equation}
\label{eq:full-equation}
i\frac{d}{dt}
\left(
 \begin{array}{c}
 {\bf a}_{m}\\
 {\bf a}_{n}
 \end{array}
\right)
=
\left(
 \begin{array}{cc}
   {\bf 0}&A  \\
  A^{\dagger}& {\bf 0}
 \end{array}
\right)
\left(
 \begin{array}{c}
 {\bf a}_{m}\\
 {\bf a}_{n}
 \end{array}
\right)
\end{equation}
where
\[
{\bf a}_{k}
=
\left(
 \begin{array}{c}
 a_{k,1}\\
 a_{k,-1}
 \end{array}
\right)\qquad \mbox{for}\quad k=m,\ n
\]
and 
\begin{eqnarray}
A\equiv A(t)&=&
\left(
 \begin{array}{cc}
  \frac{{\cal R}^{'}}{2}
  \mbox{e}^{it\left(-E_{n,1}+E_{m,1}+\Omega(m-n)\right)}& 
  -\frac{{\cal R}}{2}
  \mbox{e}^{it\left(-E_{n,-1}+E_{m,1}+\Omega(m-n)\right)} \\
  \frac{{\cal R}}{2}
  \mbox{e}^{it\left(-E_{n,1}+E_{m,-1}+\Omega(m-n)\right)}& 
  -\frac{{\cal R}^{'}}{2}
  \mbox{e}^{it\left(-E_{n,-1}+E_{m,-1}+\Omega(m-n)\right)} 
 \end{array}
\right) \nonumber \\
&=&\mbox{e}^{it\Omega(m-n)}
\left(
 \begin{array}{cc}
  \frac{{\cal R}^{'}}{2}
  \mbox{e}^{it\left(-E_{n,1}+E_{m,1}\right)}& 
  -\frac{{\cal R}}{2}
  \mbox{e}^{it\left(E_{n,1}+E_{m,1}\right)} \\
  \frac{{\cal R}}{2}
  \mbox{e}^{it\left(-E_{n,1}-E_{m,1}\right)}& 
  -\frac{{\cal R}^{'}}{2}
  \mbox{e}^{it\left(E_{n,1}-E_{m,1}\right)} 
 \end{array}
\right) 
\end{eqnarray}
because $E_{k,-\sigma}=-E_{k,\sigma}$. 

We can give (\ref{eq:full-equation}) a formal solution by infinite series 
(called Dyson series in Theoretical Physics). Then we meet secular terms. 

\par \noindent
For example let us consider the following simple equation : 
\[
\frac{d}{dt}a=\mbox{e}^{i\omega t}a \quad \mbox{with}\quad a(0)=c.
\]
The solution is given by 
\[
a(t)=
\left\{
\begin{array}{ll}
c\ \mbox{exp}\left(
\frac{\mbox{e}^{i\omega t}-1}{i\omega}
\right)\qquad \omega \ne 0 \\
c\ \mbox{e}^{t}\qquad \qquad \qquad \ \ \omega=0
\end{array}
\right.
\]
That is, we meet the secular term. 

By the way, we have known how to handle (simple) secular terms called 
Renormalization Group Method (Approach), see \cite{EFK} for a general 
introduction. 

Frasca in \cite{MFr3} has applied this method to the above equation. 
The conclusion is interesting, but seems to be rather involved. 
We are now reconsidering his approach. Therefore let us present 

\begin{flushleft}
{\bf Problem}\quad Solve this matrix equation completely ! 
\end{flushleft}
%


\end{document}